\documentclass[12pt]{article}
\usepackage{amsfonts,amssymb,amsmath}
\usepackage[dvips]{epsfig}
\usepackage[T1]{fontenc}
\usepackage[latin1]{inputenc}
\usepackage{graphicx}
\usepackage[english]{babel}
\usepackage{amsmath}
\usepackage{amssymb}
\usepackage{amsfonts}

\def\beq{\begin{equation}}
\def\eneq{\end{equation}}
\def\bea{\begin{eqnarray}}
\def\enea{\end{eqnarray}}

\begin{document}
\title{\bf{Quantum tunneling from generalized $(2+1)$ dimensional black holes
having Noether symmetry}}
\author{F. Darabi$^{1,2}$\thanks{Email: f.darabi@azaruniv.edu}, K. Atazadeh$^{1}$\thanks{Email: atazadeh@azaruniv.edu}, and A. Rezaei-Aghdam$^{1}$\thanks{Email: rezaei-a@azaruniv.edu}\\
{\small $^{1}$ Department of Physics, Azarbaijan Shahid Madani University, Tabriz 53741-161, Iran.}\\{\small $^{2}$Research Institute for Astronomy and Astrophysics of Maragha (RIAAM) Maragha 55134-441, Iran}}

\maketitle
\begin{abstract}
We have studied the Hawking radiation from {\it generalized} rotating and static $(2+1)$-dimensional BTZ black holes. In this regard, we have benefited the quantum tunneling approach with WKB approximation and obtained the tunneling rate of outgoing scalar and spinor particles across the horizons. We have also obtained the Hawking temperature at the horizons corresponding to the
emission of these particles. It is shown that the tunneling rate and Hawking temperature of generalized $(2+1)$-dimensional BTZ black holes are different from ordinary $(2+1)$-dimensional BTZ black holes due to the Noether symmetry.
In other words, the Noether symmetry can change the tunneling rate and Hawking temperature of the BTZ black holes. This symmetry may cause the BTZ black holes to avoid of evaporation and its breakdown may start the evaporation.
\end{abstract}
Pacs:{04.70.Bw; 04.50.Kd; 04.70.Dy}
\maketitle

\section{Introduction}

Hawking radiation is a quantum mechanical phenomena by which the 
(3+1)-dimensional black holes in the background of classical general relativity can evaporate \cite{Hawking:1974sw, Gibbons:1977mu}. This phenomena has also been considered as quantum tunneling of particles from the horizons of black holes \cite{Kraus:1994fh, Parikh:1999mf, Iso:2006ut}. In this approach, the Klein-Gordon or Dirak wave equations for scalar or spinor particles are solved in the spacetime background of the black holes by using complex path integration techniques and WKB approximation. This gives the tunneling rate of scalar or spinor particles across the event horizons, as well as the Hawking temperature of the black holes.

On the other hand, $(2+1)$-dimensional black holes are among the most interesting
subjects in lower dimensional gravity. The vacuum solution of $(2+1)$-dimensional gravity is flat for zero cosmological constant, hence no black hole solution exist \cite{Ida}. However, $(2+1)$-dimensional BTZ black hole solutions for a negative cosmological constant were shown to exist by Ba$\tilde{n}$ados, Teitelboim and Zanelli \cite{BTZ}. These black holes are very similar to $(3+1)$-dimensional black holes in their thermodynamical properties. Moreover, they have inner and outer horizons, mass, angular momentum and charge. Recently, we have obtained a generalized $(2+1)$-dimensional BTZ black hole solution
by using the Noether symmetry approach \cite{DAR}. (For a study of Noether
symmetry in spherical solutions, in 4D $F(R)$ and $F(T)$ gravity, see Ref.\cite{Noether}). This black hole has three conserved charges as mass $M$, angular momentum $J$ and a new conserved charge $Q_N$ corresponding respectively to the invariance of the solution under time translation, rotation, and continuous displacement of the Ricci scalar in the action. 

The quantum tunneling for scalar and spinor particles from $(2+1)$-dimensional charged and/or rotating black holes has been studied in Ref.\cite{Saif} by using WKB approximation, and the corresponding tunneling rates and Hawking temperatures have been obtained. In this paper, following Ref.\cite{Saif},
we study the Hawking radiation from the above mentioned generalized rotating and static $(2+1)$-dimensional BTZ black holes to obtain the tunneling rate of outgoing scalar and spinor particles across the horizons and the Hawking temperature at the horizons corresponding to the emission of these particles.

\section{$(2+1)$-dimensional BTZ black holes} 

The well known $(2+1)$-dimensional BTZ black hole solutions are derived from
Einstein field equations with cosmological constant in three dimensions \cite{BTZ}. The action is given by
\begin{equation}
I=\frac{1}{2\pi}\int d^{3}x\sqrt{-g}\left( R-2\Lambda \right) ,  \label{E}
\end{equation}%
where $G$ and $\Lambda =-1/{l^{2}}$ are gravitational and cosmological constants respectively, $R$ is the Ricci scalar and $g$ is determinant of the
metric tensor $g_{\mu \nu }$\footnote{We use units where $8 G=1$.}. The line element for this solution is given by
\begin{equation}
ds^{2}=-f\left( r\right) dt^{2}+f^{-1}\left( r\right) dr^{2}+r^{2}\left(
d\phi -\frac{J}{2r^{2}}dt\right) ^{2},  \label{BTZ_}
\end{equation}%
where%
\begin{equation}
f\left( r\right) =-M+\frac{r^{2}}{l^{2}}+\frac{J^{2}}{4r^{2}}.
\end{equation}%
The constants of motion as the mass $M$ and angular momentum of the BTZ black hole $J$ are appeared due to the time translation symmetry and rotational
symmetry of the metric, corresponding to the killing vectors $\partial_t$ and $\partial_{\phi}$, respectively. The Schwarzschild coordinates are defined
as
\begin{equation}
-\infty <t<\infty ,\quad 0\leq r<\infty ,\quad 0\leq \phi <2\pi .
\end{equation}%
The horizons of line element (\ref{BTZ_}) are obtained by
putting $f\left( r\right) =0$,
\begin{equation}
r_{\pm }^{2}=\frac{l^{2}}{2}\left( M\pm \sqrt{M^{2}-\frac{J^{2}}{l^{2}}}%
\right) ,
\end{equation}%
where $r_{+}$ and $r_{-}$ denote the outer and inner horizons, respectively. More general metrics have been obtained by considering coupling of the
pure Einstein gravity with other matter fields, for instance topological
matter \cite{Carlip:1991zk}, or Maxwell tensor \cite{BTZ, Martinez:1999qi}.
The action for the latter is given by
\begin{equation}
I=\frac{1}{2\pi}\int d^{3}x\sqrt{-g}\left( R-2\Lambda -\frac{\pi}{2} F_{\mu \nu
}F^{\mu \nu }\right) ,  \label{Einstein_Maxwell}
\end{equation}%
with
\begin{equation}
F_{\mu \nu }=A_{\mu ,\nu }-A_{\nu ,\mu },
\end{equation}
where $A_{\mu }$ is the electrical potential and ``,'' denotes partial derivative. The line element for electrically charged and rotating BTZ black hole solutions of Einstein-Maxwell theory is given by (\ref{BTZ_}) in which
\begin{equation}
f(r)=-M+\frac{r^{2}}{l^{2}}-\frac{1}{2}Q^{2}\ln \left( \frac{r}{l}\right) +%
\frac{J^{2}}{4r^{2}},  \label{f_r_Q__}
\end{equation}%
with $Q$ and $J$ being the electric charge and  angular momentum of the
black hole, respectively.

\section{Generalized $(2+1)$-dimensional BTZ black hole}

Recently \cite{DAR}, a generalized $(2+1)$-dimensional BTZ black hole solution has been introduced by the authors through the application of Noether symmetry in the metric formalism for $(2+1)$-dimensional $F(R)$ gravity with the following action 
\begin{equation}\label{1}
I=\frac{1}{2\pi}\int d^3x \sqrt{-g}F({R}).
\end{equation}
This action describes a theory of $(2+1)$ gravity where $F(R)$ is a typical function of the Ricci scalar $R$ subject to a Noether symmetry. 
To study the spherical solutions, one may take the metric in the following form
\begin{equation}\label{47}
ds^2=-f(r)dt^2+f^{-1}(r)dr^2+r^2[g^2(r)dt+d\phi]^2,
\end{equation}
where the radial functions $f(r)$ and $g(r)$ are regarded as the degrees of freedom. By calculating the Ricci scalar $R$, generalizing the degrees of freedom, defining a canonical Lagrangian and using the method of Lagrange multipliers to set $R$ as a constraint of the dynamics we obtain \cite{DAR}
\begin{equation}\label{6}
{\cal L}=r(F-R F_R)+\frac{r^3}{2}g'^2F_R -f' F_R +r f' F_{RR}R',
\end{equation}
where $'$ denotes the derivative with respect to $r$, $F_{R}\equiv dF/dR$
and $F_{RR}\equiv d^2F/dR^2$. Application of Noether symmetry approach
as $L_X {\cal L}=0$ results in \cite{DAR} 
\begin{equation}\label{35}
X=\left(2{A}\frac{\partial}{\partial f}+\beta_0
\frac{\partial}{\partial g}-\frac{A f'}{f}\frac{\partial}{\partial {f'}}\right),
\end{equation}
\begin{equation}\label{45}
F({R})=D_1 R\left(\frac{n}{n+1}\right)\left(\frac{R}{K}\right)^{1/n}+D_2
R+D_3,
\end{equation}
\bea\label{41}
g(r)=-{\frac {C_{{1}}{D_{{1}}}^{2}\ln  \left( D_{{1}}r+D_{{2}} \right) }{{D
_{{2}}}^{3}}}+{\frac {C_{{1}}{D_{{1}}}^{2}\ln  \left( r \right) }{{D_{
{2}}}^{3}}}
-\frac{1}{2}\,{\frac {C_{{1}}}{D_{{2}}{r}^{2}}}+{\frac {C_{{1}}D_{{
1}}}{{D_{{2}}}^{2}r}},
\enea
\begin{eqnarray}\label{46}
f(r)&=&\frac{1}{4 D_2^4 r^2}[D_2 \left(C_1^2 (6 D1 r+D_2)+8 D_2^3 r 
(r P+Q_N)\right)]\\ \nonumber
&+&\frac{1}{4 D_2^4 r^2}[2C_1^2D_1r\ln\frac{r}{D_1r+D_2}(2D_2+3D_1r)]-\frac{K r^{n+2}}{n^2+5 n+6},
\end{eqnarray}
where $A, \beta_0, D_1, D_2, D_3, C_1, K, P$ are constants, $Q_N$ is the
Noether charge due to the Noether symmetry and $n$ can take arbitrary values. Note that as a direct consequence of imposing the Noether symmetry the constant $D_3$ does not appear in the solution for $f(r)$. Therefore, the solutions $f(r)$ and $g(r)$ have symmetry under a change in the value of $D_3$. 

Let us now consider a very simple example $F(R)=R+D_3$ associated to the special case $D_1=0, D_2=1$. Then, we obtain 
\begin{equation}\label{49'}
f(r)=\frac{C_2^2}{4r^2}-\frac{K}{n^2+5 n+6}r^{n+2}+\frac{2Q_N}{r}+2P.
\end{equation}
Assuming $n=0$ and using the identifications $2P=-M$, $C_2=J$, and $K=-6l^{-2}$
we obtain  
\begin{equation}\label{52'}
f(r)=-M-\frac{R}{6}r^{2}+\frac{2Q_N}{r}+\frac{J^2}{4r^2},
\end{equation}
\begin{equation}\label{52''}
g(r)=-\frac{J}{2r^2}.
\end{equation}
Note that unlike $M$ and $J$ which are related to the killing vectors preserving the metric, the Noether charge $Q_N$ is not related to such a killing vector, rather it results due to the Killing vector which is related to the Noether symmetry of the action. For the simple example $F(R)=R+D_3$ with the specific Noether symmetry
subject to the Killing vector $X$, we may find the Noether charge as \cite{DAR}
\bea\label{charge}
Q_N=-2A(F_R+r F_{RR}R')+\beta_0r^3 g' F_R =
-2A+\beta_0 J.    \label{1.9}
\enea
Note that the Noether charge is independent of the mass $M$, but depends on the angular momentum $J$. For non-rotating black hole with $J=0$, the Noether charge reduces to $Q_N=-2A$. In the following, we will study the quantum tunneling of scalar and spinor particles from generalized $(2+1)$-dimensional BTZ black holes having mass $M$, angular momentum $J$ and Noether charge $Q_N$ given by \eqref{charge}.

\section{Quantum tunneling of scalar particles from generalized $(2+1)$-dimensional
BTZ black holes}

\label{sec_ charge_scalar_charge}

The emission of scalar particles from $(2+1)$-dimensional black holes
may be considered as a tunneling phenomenon across their event horizons \cite{Saif}. To investigate this phenomenon in the case of (electrically charge-less) generalized BTZ black hole given by (\ref{52'}) and (\ref{52''}), we should solve the Klein-Gordon equation for the neutral massive
scalar field, $\Psi $, which is given by
\begin{equation}
\frac{1}{\sqrt{-g}}\partial _{\mu }
\left( \sqrt{-g}g^{\mu \nu }\partial _{\nu }\Psi
\right) -\frac{m^{2}}{\hbar ^{2}}\Psi =0,  \label{K_G_1}
\end{equation}%
where $\mu $, $\nu$=0,1,2 corresponding to the coordinates $t,r,\phi$
and $m$ is the mass of the particle. Using WKB approximation and taking an ansatz of the form \cite{Saif}
\begin{equation}
\Psi (t,r,\phi )=e^{\left( \frac{i}{\hslash }I(t,r,\phi )+I_{1}(t,r,\phi
)+O(\hslash )\right) },  \label{wave_factor_}
\end{equation}%
the scalar field equation (\ref{K_G_1}), in leading powers of $\hbar$, results
in 
\begin{equation}
g^{tt}(\partial _{t}I)^{2}+g^{rr}(\partial _{r}I)^{2}+g^{t\phi
}(\partial _{t}I\partial _{\phi }I)+g^{\phi \phi
}(\partial _{\phi }I)^{2}+m^{2}=0.
\end{equation}%
or
\begin{eqnarray}
-f(r)^{-1}(\partial _{t}I)^{2}+ f(r) (\partial
_{r}I)^{2}+r^{-2}g(r)^{-2}(\partial _{t}I\partial _{\phi }I)+r^{-2}(\partial _{\phi }I)^{2}+m^{2}=0,  \label{wave_Q_}
\end{eqnarray}
where $f(r)$ and $g(r)$ are given by (\ref{52'}) and (\ref{52''}). Regarding the killing vector fields $\partial _{t}$ and $\partial _{\phi }$ which preserve the metric, one may find that the above differential equation has a solution which can be written in terms of the classical action $I$ given by
\begin{equation}
I=-\omega t+W(r)+j\phi +K,  \label{phase_}
\end{equation}%
where $\omega$ and $j$ denote the energy and angular momentum of the particle
respectively, and $K$ is a constant which can be complex. Inserting $I$ in the differential equation (\ref{wave_Q_}) we obtain
\begin{equation}
W_{\pm }(r)=\pm \int \frac{\sqrt{\omega^{2}-(\frac{f(r)}{r^{2}}%
- g(r)^{2})j^{2}+2g(r)\omega j-f(r)m^{2}}%
}{f(r)}dr.
\end{equation}%
Since $f(r_{+})=0$, by using the residue theory for semi circles we have simple pole at $r=r_{+},$ so we get
\begin{equation}\label{Wpm}
W_{\pm }=\pm i\pi \frac{ \omega +j g(r_{+})}{f^{\prime}(r_{+})}.
\end{equation}
This implies that
\begin{equation}
\mathrm{Im}W_{+}=\pi \frac{ \omega +j g(r_{+}) }{f^{\prime }(r_{+})}.  \label{phase_imaginary_Q}
\end{equation}

We know that Hawking radiation from black holes may be considered as a process
of quantum tunneling by which the particles can tunnel across the black hole horizon from inside to the outside. Using this viewpoint to calculate the rate of tunneling, we may resort to the semi-classical method and calculate the imaginary part of the classical action. In this regard, the tunnelling
rates across the horizon from inside to outside $\Gamma_{em}$, and from outside to inside $\Gamma_{ab}$, are given respectively by \cite{Shankaranarayanan:2000gb,
Srinivasan:1998ty}
\begin{eqnarray}
\Gamma_{em} &= &\exp \left( \frac{-2}{\hbar }\mathrm{Im}I\right)
=\exp \left( \frac{-2}{\hbar }(\mathrm{Im}W_{+}+\mathrm{Im}K)\right) ,
\label{p_emit_} \\
\Gamma_{ab} &= &\exp \left( \frac{-2}{\hbar }\mathrm{Im}I\right)
=\exp \left( \frac{-2}{\hbar }(\mathrm{Im}W_{-}+\mathrm{Im}K)\right) .
\label{p_absorb_}
\end{eqnarray}%
Since any outside particle will certainly fall inside the black hole,
we must have $\Gamma_{ab}=1$ which results in Im$K=-$Im$W_{-}$. On the other
hand, from (\ref{Wpm}) we have $W_{+}=-W_{-}$ which means that the tunneling rate of a particle from inside to the outside of the horizon is
\begin{equation}
\Gamma_{em} =\exp \left( \frac{-4}{\hbar }\mathrm{Im}W_{+}\right) .
\label{rate_phase_}
\end{equation}%
Putting Im$W_{+}$ from equation (\ref{phase_imaginary_Q}) into (%
\ref{rate_phase_}) results in
\begin{equation}
\Gamma_{em} =\exp \left( \frac{-4\pi(\omega+j g(r_{+}))}{\hbar
f^{\prime }(r_{+})}\right),  \label{rate_}
\end{equation}%
or
\begin{equation}
\Gamma_{em} =\exp \left( \frac{-4\pi(\omega-j \frac{J}{2r_+^2})}{\hbar
\left( -\frac{Rr_{+}}{3}-\frac{2Q_{N}}{r_{+}^2}-\frac{J^2}{2r_{+}^3}\right)}\right),.  \label{rate_Q}
\end{equation}%
where $Q_N=-2A+\beta_0 J$.
This is the tunneling rate of scalar particles from inside to the outside
of the event horizon of the generalized $(2+1)$-dimensional BTZ black hole.

Comparing Eq.(\ref{rate_Q}) with the Boltzmann factor $\Gamma=\exp
\left( -\beta \omega \right) $, where $\omega$ and $\beta$ denote the energy of particle and the inverse temperature of the horizon, respectively \cite%
{Shankaranarayanan:2000gb, Srinivasan:1998ty}, we can derive the Hawking
temperature as ($\hbar=1$)
\begin{equation}
T_{H}=\frac{f^{\prime }(r_{+})}{4\pi }, \label{T_tunneling_01}
\end{equation}%
which results in
\begin{equation}
T_{H}=\frac{1}{4\pi }\left(-\frac{Rr_{+}}{3}-\frac{2Q_{N}}{r_{+}^2}-\frac{J^2}{2r_{+}^3}\right),  \label{T_tunneling_Q}
\end{equation}%
where $Q_N=-2A+\beta_0 J$. 
The tunneling rate and Hawking temperature corresponding to a static ($J=0$) generalized BTZ black hole is given respectively by
\begin{equation}
\Gamma_{em} =\exp \left( \frac{-4\pi\omega}{\hbar
\left(-\frac{Rr_{+}}{3}-\frac{2Q_{N}}{r_{+}^2}\right)}\right).  \label{rate_Q'}
\end{equation}%
\begin{equation}
T_{H}=\frac{1}{4\pi }\left(-\frac{Rr_{+}}{3}-\frac{2Q_{N}}{r_{+}^2}\right)
,  \label{T_tunneling_Q'}
\end{equation}%
where $Q_N=-2A$. 
Considering (\ref{rate_Q}), $(\ref{T_tunneling_Q})$,
for rotating black hole, and (\ref{rate_Q'}), (\ref{T_tunneling_Q'}),
for static black hole, we realize that the tunneling rates and Hawking temperatures vanish provided that
\begin{equation}
\omega-j \frac{J}{2r_+^2}>0\,\,\,\,, \,\,\,\, -\frac{Rr_{+}}{3}-\frac{2Q_{N}}{r_{+}^2}-\frac{J^2}{2r_{+}^3}=0.
\label{T_tunneling_J}
\end{equation}%
and
\begin{equation}
\omega>0\,\,\,\,, \,\,\,\, -\frac{Rr_{+}}{3}-\frac{2Q_{N}}{r_{+}^2}=0,
\label{T_tunneling_0}
\end{equation}%
respectively. Now, the Noether charge plays its key role. Unlike the other constants, namely mass $M$ and angular momentum $J$, the Noether charge $Q_{N}$ is not an intrinsic feature of the black hole, because it is related to the freedom in taking the arbitrary values for the Ricci scalar. Hence, the generalized $(2+1)$-dimensional BTZ black hole has a free extrinsic parameter $Q_{N}$ which can be tuned so as to set the tunneling rate and Hawking temperature zero. This is remarkable in that the generalized $(2+1)$-dimensional black
holes may avoid of evaporation due to the existence of the Noether symmetry.

\section{Quantum tunneling of fermionic particles from generalized $(2+1)$-dimensional
black holes}

In this section, we will deal with fermionic particles and calculate the
corresponding tunneling rate and Hawking temperature for the generalized $(2+1)$-dimensional BTZ black holes. In this regard, we should solve the $(2+1)$-dimensional Dirac equation for the two-component charge-less spinor field $\psi $ with the mass $m$ as
\begin{equation}
i{\hbar }\gamma ^{\mu }\left( \partial _{\mu }+\Omega _{\mu }\right) \psi -{m}\psi =0,
\end{equation}
where
\begin{eqnarray}
\Omega _{\mu } &=&\frac{i}{2}\Gamma _{\mu }^{\alpha \beta }\Sigma _{\alpha
\beta }, \\
\Sigma _{\alpha \beta } &=&\frac{i}{4}\left[ \gamma ^{\alpha },\gamma
^{\beta }\right] ,~~~~\Omega _{\mu }=\frac{-1}{8}\Gamma _{\mu }^{\alpha
\beta }\left[ \gamma ^{\alpha },\gamma ^{\beta }\right] .
\end{eqnarray}
Using the Pauli matrices $\sigma ^{i}$ as
\begin{equation}
\sigma ^{0}=\left(
\begin{array}{cc}
0 & 1 \\
1 & 0%
\end{array}%
\right) ,\,\,\sigma ^{1}=\left(
\begin{array}{cc}
0 & -i \\
i & 0%
\end{array}%
\right) ,\,\,\sigma ^{2}=\left(
\begin{array}{cc}
1 & 0 \\
0 & -1%
\end{array}%
\right),
\end{equation}%
we take the curved space $\gamma $ matrices in $(2+1)$-dimensional spacetime as
\begin{eqnarray}
\gamma ^{t} &=&\left(
\begin{array}{cc}
0 & -\frac{1}{\sqrt{f}} \\
\frac{1}{\sqrt{f}} & 0%
\end{array}%
\right) ,\nonumber \\
\gamma ^{r}&=&\left(
\begin{array}{cc}
0 & \sqrt{f} \\
\sqrt{f} & 0%
\end{array}%
\right) ,  \nonumber \\
\gamma ^{\phi }&=&\left(
\begin{array}{cc}
\frac{1}{r} & -\frac{J}{2r^{2}\sqrt{f}} \\
\frac{J}{2r^{2}\sqrt{f}} & -\frac{1}{r}%
\end{array}%
\right) ,
\end{eqnarray}%
which satisfy the condition $\left\{ \gamma ^{\mu },\gamma ^{\nu
}\right\} =2g^{\mu \nu }$. Then, Dirac equation takes the form
\begin{equation}
i\gamma ^{t} \partial _{t} \psi +i\gamma
^{r}\left( \partial _{r}\right) \psi +i\gamma ^{\phi }\left( \partial _{\phi
}\right) \psi -\frac{\mu }{\hbar }\psi =0.  \label{Dirac_}
\end{equation}%
Considering the fact that the wave function for a fermion with spin 1/2 has two states namely spin-up ($\uparrow $) and spin-down ($\downarrow $), we take the following ansatz respectively for the solution \cite{Saif}
\begin{eqnarray}
\psi _{\uparrow } &=&\left(
\begin{array}{c}
A\left( t,r,\phi \right)  \\
0%
\end{array}%
\right) e^{\frac{i}{\hslash }I_{\uparrow }\left( t,r,\phi \right) },
\label{spin_up_} \\
\psi _{\downarrow } &=&\left(
\begin{array}{c}
0 \\
B\left( t,r,\phi \right)
\end{array}%
\right) e^{\frac{i}{\hslash }I_{\downarrow }\left( t,r,\phi \right) }.
\label{spin_down_}
\end{eqnarray}%
Inserting (\ref{spin_up_}) for the spin-up particle into the Dirac equation
(\ref{Dirac_}) and after a simple manipulation, we obtain the following equation
\begin{equation}
-\frac{A}{\sqrt{f}}\partial _{t}I_{\uparrow }\left( t,r,\phi \right) -\sqrt{f}A\partial _{r}I_{\uparrow }\left( t,r,\phi
\right) -\frac{JA}{2r^{2}\sqrt{f}}\partial _{\phi }I_{\uparrow }\left(
t,r,\phi \right) =0.
\end{equation}%
Using the method of separation of variables for the spin-up state we have
\begin{equation}
I_{\uparrow }=-\omega t+W(r)+\Theta (\phi )+K=-\omega t+W(r)+j\phi +K,
\label{spin_up_phase}
\end{equation}%
where, as before, $\omega $ and $j$ are the energy and angular momentum of the emitted particle, and $K$ can be a complex constant. Putting this expression in the above equation we obtain
\begin{equation}
\frac{A}{\sqrt{f}}\omega -\sqrt{f}A\partial _{r}W-%
\frac{JA}{2r^{2}\sqrt{f}}\partial _{\phi }\Theta =0.  \label{eqn_spin_}
\end{equation}%
Considering $\Theta (\phi )=j\phi$ and removing $A$, we obtain
\begin{equation}
\frac{\omega }{\sqrt{f}}+\frac{qA_{t}}{\sqrt{f}}-\sqrt{f}\partial _{r}W-j%
\frac{J}{2r^{2}\sqrt{f}}=0,
\end{equation}%
or
\begin{equation}
\partial _{r}W=\frac{1}{f(r)}\left( \omega -j\frac{J}{2r^{2}}\right) .
\label{spin_up_w}
\end{equation}
Integration of equation (\ref{spin_up_w}) along a semi circle around the pole at $r_{+}=0$ results in the radial function at the horizon as
\begin{equation}
W(r_+)=\frac{\pi i(\omega -j\frac{J}{2r_{+}^{2}})}{\left(-\frac{Rr_{+}}{3}-\frac{2Q_{N}}{r_{+}^2}-\frac{J^2}{2r_{+}^3}\right) }.
\end{equation}%
Using equation (\ref{rate_phase_}), the tunneling rate is given by  
\begin{equation}
\Gamma_{em} =\exp \left( \frac{-4\pi (\omega -j\frac{J}{2r_{+}^{2}})}{\hbar\left( -\frac{Rr_{+}}{3}-\frac{2Q_{N}}{r_{+}^2}-\frac{J^2}{2r_{+}^3}\right)}\right),
\label{tunneling_c_r_f}
\end{equation}%
where $Q_N=-2A+\beta_0 J$.
We find that the tunneling rate for a fermionic particle across the horizon
of the generalized rotating BTZ black hole is the same as that of obtained for the scalar field in (\ref{rate_Q}). It is then obvious that the temperature of this black hole at its horizon, regarding the emission of fermionic particles,
is given by the expression (\ref{T_tunneling_Q}). Accordingly, the tunneling
rate and temperature of the generalized static ($J=0$) BTZ black hole are
given by the expressions (\ref{rate_Q'}) and (\ref{T_tunneling_Q'}), respectively.

\section{Conclusions}

In this paper, we have used the quantum tunneling approach and WKB approximation
to calculate the tunneling rate of outgoing scalar and spinor particles across the horizons of rotating and static generalized $(2+1)$-dimensional BTZ black holes. We have also calculated the Hawking temperature at the horizons corresponding to the emission of these particles. We have shown that the generalized $(2+1)$-dimensional BTZ black holes, with an extrinsic parameter $Q_{N}$ as the Noether charge,
have different tunneling rate and Hawking temperature from those of ordinary
$(2+1)$-dimensional BTZ black holes. In other words, the Noether symmetry
can change the tunneling rate and Hawking temperature of the BTZ black holes. This is remarkable, because the generalization of ordinary $(2+1)$-dimensional BTZ black holes by applying Noether symmetry may cause these black holes
to avoid of evaporation. When the free parameter $D_3$ in the generalized action is fixed to a cosmological constant $\Lambda=-l^{-2}$ as $D_3=2l^{-2}$, \, the Noether symmetry is broken, namely $Q_N=0$, and one recovers the ordinary
BTZ black holes having $R=-6l^{-2}$. Then, the tunneling rate and Hawking temperature of the generalized $(2+1)$-dimensional BTZ black holes coincide with those of ordinary $(2+1)$-dimensional BTZ black holes and they start evaporation.

\section*{Acknowledgment}
This work has been supported financially by Research Institute
for Astronomy and Astrophysics of Maragha (RIAAM) under research project NO.1/3252-44.

\end{document}